\documentclass[twocolumn,letterpaper,prb, floats]{revtex4}
\usepackage{color,graphicx,url}
\usepackage{amsmath,amssymb,amsfonts,amsthm,subfigure}
\usepackage[colorlinks, urlcolor=black, citecolor=black, linkcolor=black]{hyperref}

\usepackage{algorithm}
\usepackage{algpseudocode}

\usepackage{ifpdf}
\ifpdf
\fi
\date{\today}
\begin{document}

\title{Non-monotonic recursive polynomial expansions for linear scaling
  calculation of the density matrix }
\begin{abstract}
As it stands, density matrix purification is a powerful tool for
linear scaling electronic structure calculations.  The convergence is
rapid and depends only weakly on the band gap.
However, as will be shown in this paper, there is room for
improvements. 
The key is to allow for non-monotonicity in the recursive polynomial
expansion.
Based on this idea, new purification schemes are proposed that require
only half the number of matrix-matrix multiplications compared to
previous schemes.
The speedup is essentially independent of the location of the chemical
potential and increases with decreasing band gap.

\end{abstract}  
\author{Emanuel~H.~Rubensson}
\email{emanuel.rubensson@it.uu.se}
  \affiliation{\mbox{Division of Scientific Computing,}
    \mbox{Department of Information Technology,}
    \mbox{Uppsala University,}
    Box 337,
    SE-751 05 Uppsala,
    Sweden}
  
  \maketitle

  During the last two decades, methods have been developed that make
  it possible to apply \emph{ab initio} electronic structure
  calculations, using Hartree-Fock, Kohn-Sham density functional
  theory, or tight-binding models, to systems with many thousands of
  atoms.\cite{linsca-g, linsca-bmg, saad-review2010,
    Hine-ONETEP-20091041, linmemDFT} Although the computational cost
  of these methods increases only linearly with system size, such
  calculations are extremely demanding.  Therefore, there is a need to
  improve existing linear scaling methods in order to reduce the
  computational cost and make best use of modern computer resources.

  In linear scaling electronic structure calculations, efficient
  computation of the one-particle density matrix $D$ for a given
  effective Hamiltonian $F$ is an important ingredient.
  Many methods for linear scaling computation of the density matrix
  have been proposed.  A common approach is to employ a polynomial
  expansion of the function $D = \theta(\mu I - F)$, where $\theta$ is the
  Heaviside step function and $\mu$ is the chemical potential. The
  expansion may be built up serially by a Chebyshev
  series\cite{foe-gc, foe-gt, foe-bh, foe-lssbbh} or recursively by
  density matrix purification\cite{pur-pm, pur-n, pur-ntc, pur-h,
    pur-m} or sign matrix methods.\cite{sign-bcm, sign-ns} Another
  approach is to minimize an energy functional with respect to the
  density matrix.\cite{dmm-lnv, dmm-hp, dmm-hloj, dmm-sshw}

  For the isolated problem of computing the density matrix for a fixed
  Hamiltonian, the recursive density matrix purification schemes are
  highly efficient. The convergence is rapid and the computational
  cost scales as $\mathcal{O}(\ln (\Delta\epsilon / \xi))$ where
  $\Delta\epsilon$ is the spectral width of the effective Hamiltonian
  matrix and $\xi$ is the band gap.\cite{pur-n, puri_vs_lnv} This
  should be compared to an $\mathcal{O}(\sqrt{\Delta\epsilon / \xi})$
  cost for the serial polynomial expansion\cite{foe-lssbbh} and
  minimization\cite{linsca-g, puri_vs_lnv} methods.  However, despite
  the excellent performance of previously proposed density matrix
  purification schemes, substantial improvements are still possible as
  will be shown in this letter.

  In density matrix purification, the effective Hamiltonian matrix is
  first shifted and scaled so that the eigenvalues end up in the
  $[0,\ 1]$ interval in reverse order. After that, low order
  polynomials with fixed points at $0$ and $1$ are recursively applied
  to build up the desired step function. The general iterative
  procedure can be formulated as
  \begin{equation} \label{eq:puri_general}
    \begin{array}{rcl}
      X_0 & = & f_0(F) \\
      X_i & = & f_i(X_{i-1}),\quad i=1,2,\dots 
    \end{array}
  \end{equation}
  where $f_0$ is the initial linear transformation and $f_i,
  i=1,2,\dots$ is a sequence of low order polynomials.

  Purification can either be carried out with fixed or varying
  chemical potential $\mu$. In case of fixed-$\mu$ purification, a
  single polynomial with an unstable fixed point in $]0,\ 1[$ is
  typically used for all $f_i, i>0$. The initial transformation $f_0$
  maps the chemical potential to the unstable fixed point. The
  purification process then brings the eigenvalues to their desired
  values of $0$ and $1$.  In case of varying-$\mu$ purification, the
  chemical potential is allowed to move during the iterations. This
  flexibility can be used to automatically adjust the expansion so
  that the correct number of electrons is obtained, as in
  canonical\cite{pur-pm} and trace-correcting\cite{pur-n}
  purification.
   
  In any case, the idea has been to use polynomials that increase
  monotonically in $[0, \ 1]$ and have fixed points and vanishing
  derivatives at $0$ and $1$.
  As discussed by Niklasson,\cite{pur-n} it can be understood that a
  recursive expansion using such polynomials will converge towards a
  step function.  In the following, we shall use the notation
  $P_{i,j}(x)$ for the polynomial of degree $1+i+j$ with fixed points
  at $0$ and $1$ and with $i$ and $j$ vanishing derivatives at $0$ and
  $1$, respectively.  Many previously proposed purification
  polynomials can be written in this form.\cite{mhnm-polys}

  In this letter, we withdraw from the idea of using monotonically
  increasing purification polynomials. A scale and fold technique
  giving non-monotonic purification transformations is proposed that
  results in improved performance of both fixed- and varying-$\mu$
  purification schemes.
  The new idea is the following -- before each iteration, the
  eigenspectrum is stretched out outside the $[0,\ 1]$ interval. Some
  of the polynomials of the form $P_{i,j}$ can then be used to fold
  the eigenspectrum over itself.  For example, the polynomial
  $P_{1,0}(x) = x^2$ can be used to fold the unoccupied part of the
  eigenspectrum if the eigenspectrum is stretched out below $0$ before
  its application. Similarly, the polynomial $P_{0,1}(x) = 2x-x^2$ can
  be used to fold the occupied part.  In general, the scale and fold
  technique can for a polynomial $P_{i,j}$ be used for the unoccupied
  part if $i$ is odd and for the occupied part if $j$ is odd.

  Similar scaling techniques have previously been employed to improve
  the convergence of Newton iterations for sign matrix
  evaluations.\cite{kenney_laub_1992, book-higham} However, in this
  case the regular unscaled iteration keeps the eigenvalues outside
  the interval and the scaling is used to shrink rather than stretch
  out the eigenspectrum.

  \begin{figure*}%[h]
    \center
    \subfigure[$\, $ First iteration\label{fig:mapping_1_mcw}]{
      \includegraphics[width=0.3\textwidth]{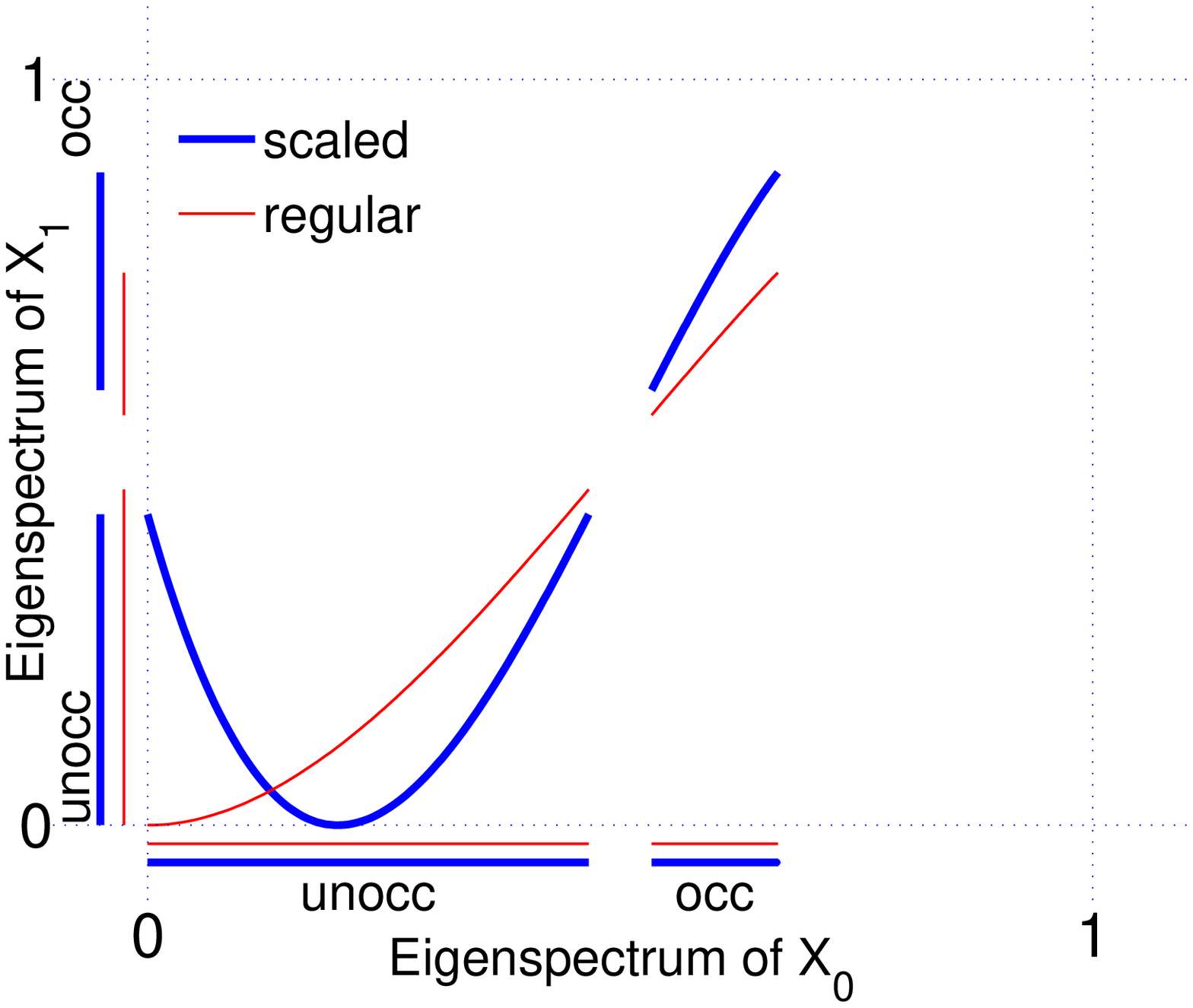}
    }
    \subfigure[$\, $ Third iteration\label{fig:mapping_3_mcw}]{
      \includegraphics[width=0.3\textwidth]{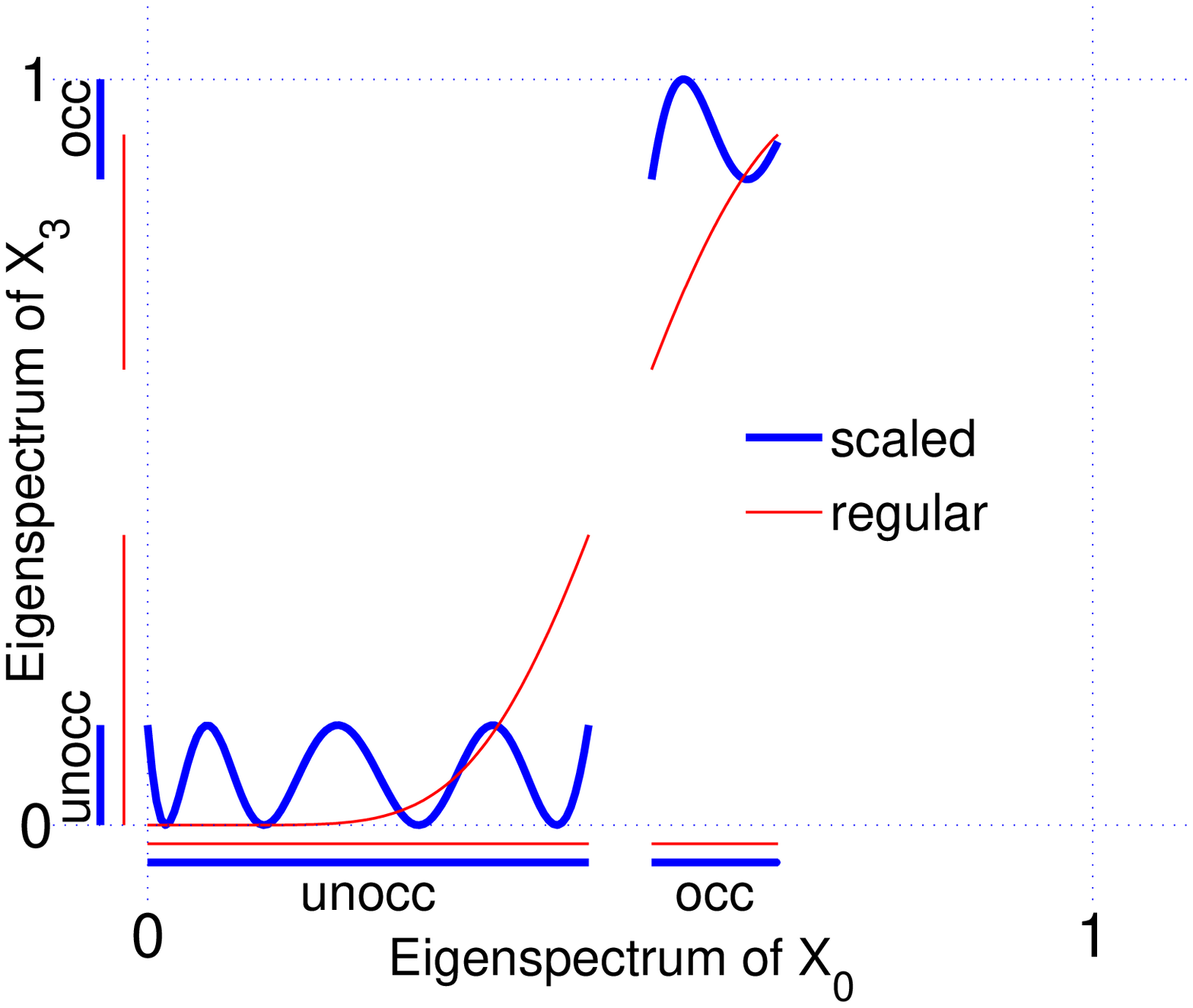}
    }
    \subfigure[$\, $ Fifth iteration\label{fig:mapping_5_mcw}]{
      \includegraphics[width=0.3\textwidth]{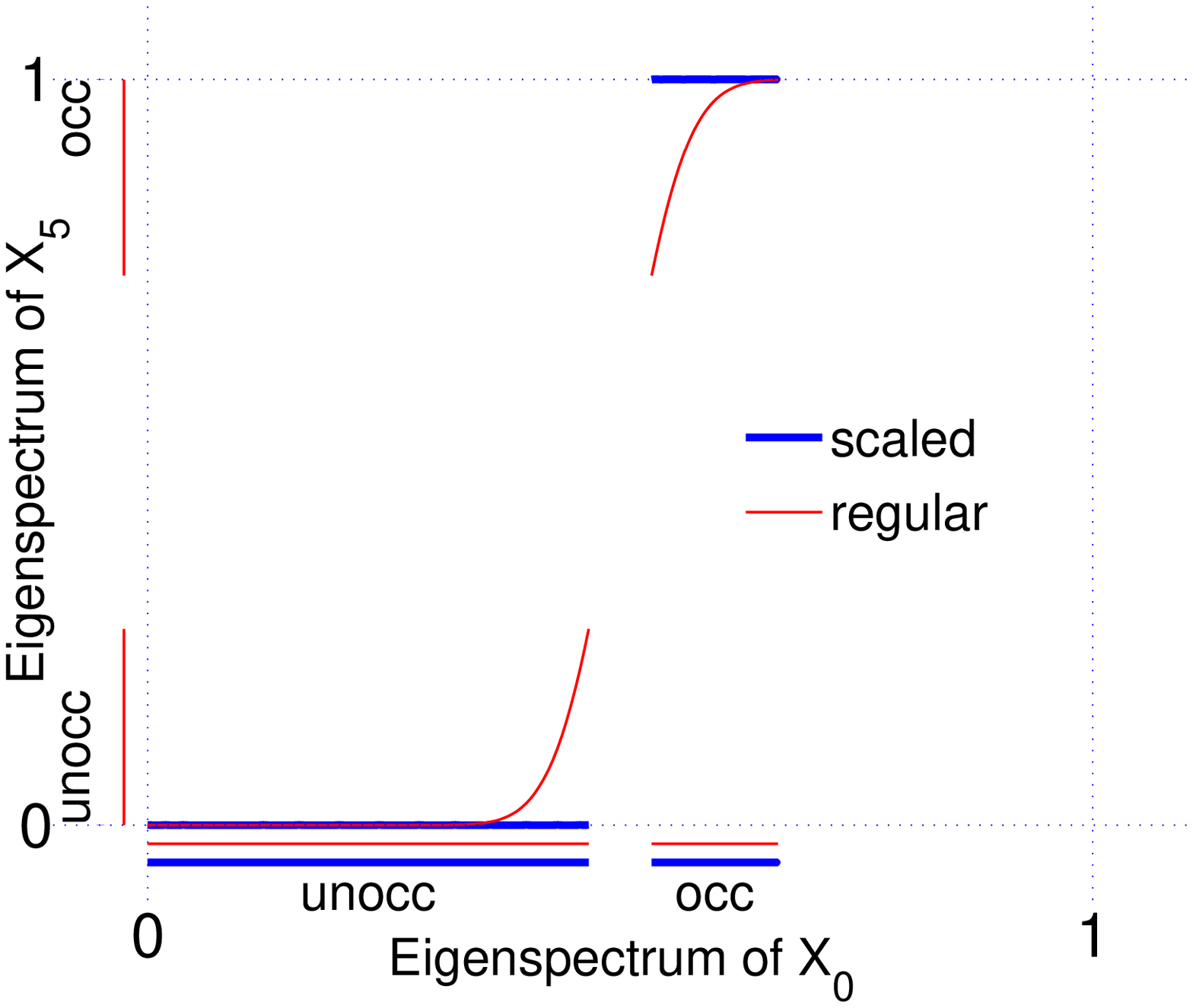}
    }
    \caption{Mapping of the eigenspectrum after 1, 3, and 5 iterations
      respectively of McWeeny based fixed-$\mu$ purification with and
      without use of scaling. In this illustrative example
      $\Delta\epsilon / \xi = 10$ and the chemical potential $\mu$ is
      located at $\lambda_{\textrm{min}} + 0.25
      (\lambda_{\textrm{max}} -
      \lambda_{\textrm{min}})$. \label{fig:mapping_mcw}}
  \end{figure*}

  We will first apply the scale and fold technique to fixed-$\mu$
  purification using a polynomial $P_{m,m}$ with $m$ being odd. For
  such polynomials, the technique can be used to fold both the
  unoccupied and occupied parts of the eigenspectrum in each
  iteration.  In this case the non-monotonic purification
  transformation
  \begin{equation}
    f_i(X_{i-1}) = P_{m,m}(\alpha(X_{i-1} - 0.5) + 0.5I)
  \end{equation}
  where $\alpha \geq 1$ determines the amount of scaling around 0.5.
  The complete algorithm for the special case $m=1$ is given in
  Algorithm~\ref{alg:fixed_mu_puri_mcw}, where
  $\lambda_{\textrm{min}}$ and $\lambda_{\textrm{max}}$ are the
  extremal eigenvalues of $F$ or bounds thereof. For simplicity, it is
  assumed here that the band gap is located symmetrically around
  $\mu$. The expression for $\alpha$ can be derived by solving
  \begin{equation}
    P_{m,m}(\alpha(\beta-0.5)+0.5) = P_{m,m}(0.5(1-\alpha))
  \end{equation}
  for $\alpha \geq 1$.  Here, $\beta$ is a parameter depending on the
  eigenvalue closest to 0.5, see
  Algorithm~\ref{alg:fixed_mu_puri_mcw}.  The behavior of
  Algorithm~\ref{alg:fixed_mu_puri_mcw} is illustrated in
  Figure~\ref{fig:mapping_mcw}. The behavior of the regular
  grand-canonical purification algorithm,\cite{pur-pm} corresponding
  to Algorithm~\ref{alg:fixed_mu_puri_mcw} with $\alpha = 1$, is shown
  for reference.
  Note how the scaled variant is able to take advantage of the
  additional flexibility given by allowing for non-monotonicity,
  resulting in much faster convergence.
  Fixed-$\mu$ purification schemes with scaling can also be derived
  for other polynomials of the form $P_{i,j}$ where $i$ and $j$ are
  both odd and larger than 0.  Note that the scaling should be
  performed around the unstable fixed point of the polynomial which
  will differ from 0.5 if $i \neq j$.

  \begin{algorithm}
    \caption{McWeeny based fixed-$\mu$ purification}
    \label{alg:fixed_mu_puri_mcw}
    \begin{algorithmic}[1]
      \Require $F,\lambda_{\textrm{min}},\lambda_{\textrm{max}},\mu,\xi$ 
      \State $\gamma = 2\max(\lambda_{\textrm{max}}-\mu, \mu-\lambda_{\textrm{min}})$
      \State $X_0 = \frac{\mu I - F}{\gamma} + 0.5 I$
      \State $\beta = 0.5(1 - \xi/\gamma)$
      \For{ $i=1,2,\dots,n$ } 
      \State $\alpha = 3 / \sqrt{ 12\beta^2-18\beta+9}$
      \State $X_s = \alpha(X_{i-1}-0.5 I)+0.5 I$
      \State $X_i = 3X_s^2-2X_s^3$
      \State $\beta_s = \alpha(\beta-0.5)+0.5$
      \State $\beta = 3\beta_s^2-2\beta_s^3$
      \EndFor
      \State \textbf{return} $D = X_n$
    \end{algorithmic}
  \end{algorithm}

  \begin{algorithm}
    \caption{$P_{0,1}$ \& $P_{1,0}$ based varying-$\mu$ purification}
    \label{alg:varying_mu_puri_tc2}
    \begin{algorithmic}[1]
      \Require $F,\lambda_{\textrm{min}},\lambda_{\textrm{max}},\lambda_{\textrm{lumo}}, \lambda_{\textrm{homo}}$ 
      \State $X_0 = f_0(F) = (\lambda_{\textrm{max}} I -
      F)/(\lambda_{\textrm{max}}-\lambda_{\textrm{min}})$
      \State $\beta = f_0(\lambda_{\textrm{lumo}})$
      \State $\bar{\beta} = f_0(\lambda_{\textrm{homo}})$
      \For{ $i=1,2,\dots,n$ } 
      \If {$\beta + \bar{\beta} > 1$}
      \State $\alpha = \beta / (2-\beta)$
      \State $X_i = ((1+\alpha)X_{i-1}-\alpha I)^2$
      \State $\beta = ((1+\alpha)\beta-\alpha)^2$
      \State $\bar{\beta} = ((1+\alpha)\bar{\beta}-\alpha)^2$
      \Else
      \State $\alpha = (1-\bar{\beta}) / (1+\bar{\beta})$
      \State $X_i = 2(1+\alpha)X_{i-1}-(1+\alpha)^2X_{i-1}^2$
      \State $\beta = 2(1+\alpha)\beta-(1+\alpha)^2\beta^2$
      \State $\bar{\beta} = 2(1+\alpha)\bar{\beta}-(1+\alpha)^2\bar{\beta}^2$
      \EndIf
      \EndFor
      \State \textbf{return} $D = X_n$
    \end{algorithmic}
  \end{algorithm}

  \begin{figure*}%[h]
    \center
    \subfigure[$\, $ First iteration\label{fig:mapping_1_tc2}]{
      \includegraphics[width=0.3\textwidth]{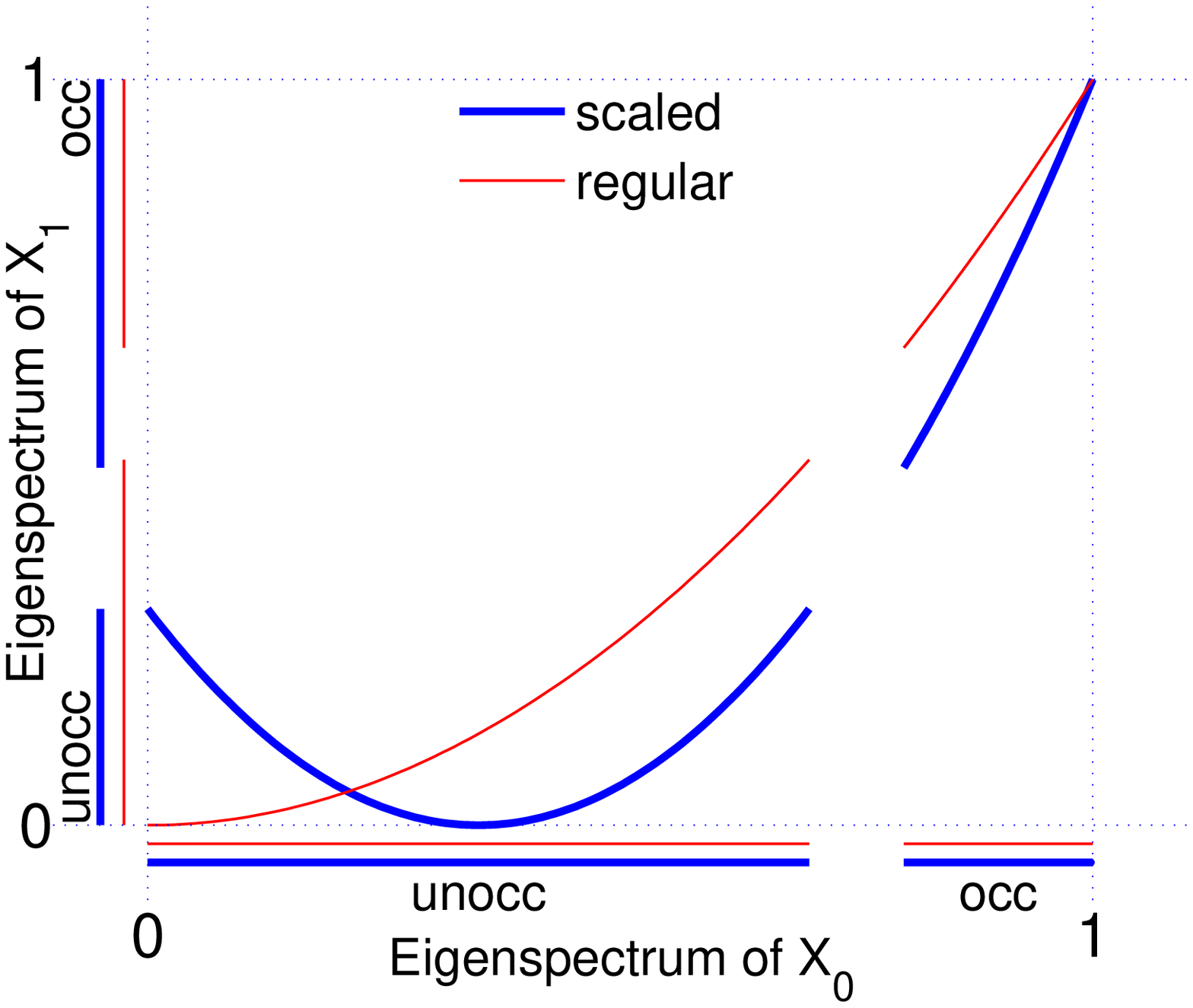}
    }
    \subfigure[$\, $ Fifth iteration\label{fig:mapping_5_tc2}]{
      \includegraphics[width=0.3\textwidth]{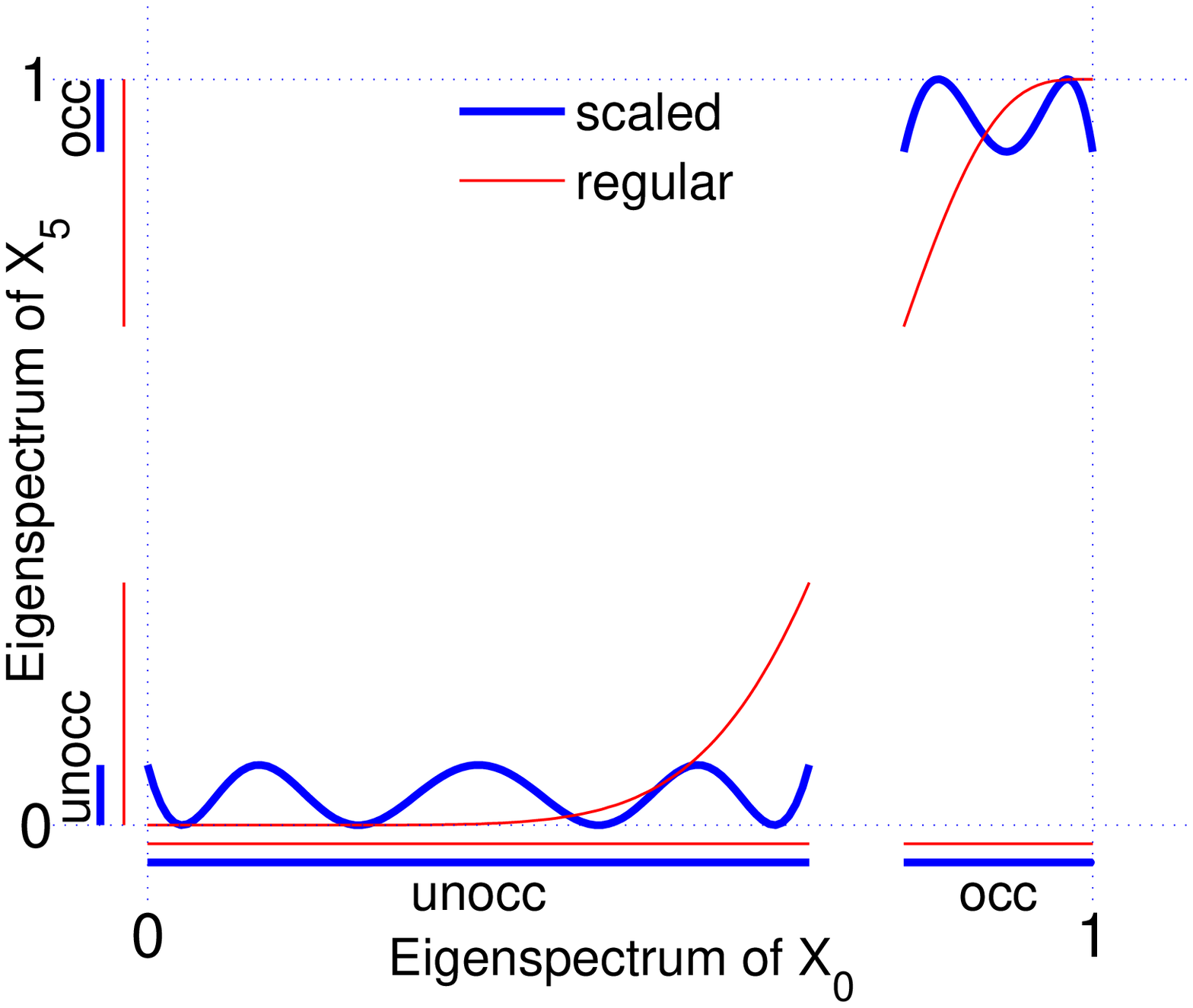}
    }
    \subfigure[$\, $ Ninth iteration\label{fig:mapping_9_tc2}]{
      \includegraphics[width=0.3\textwidth]{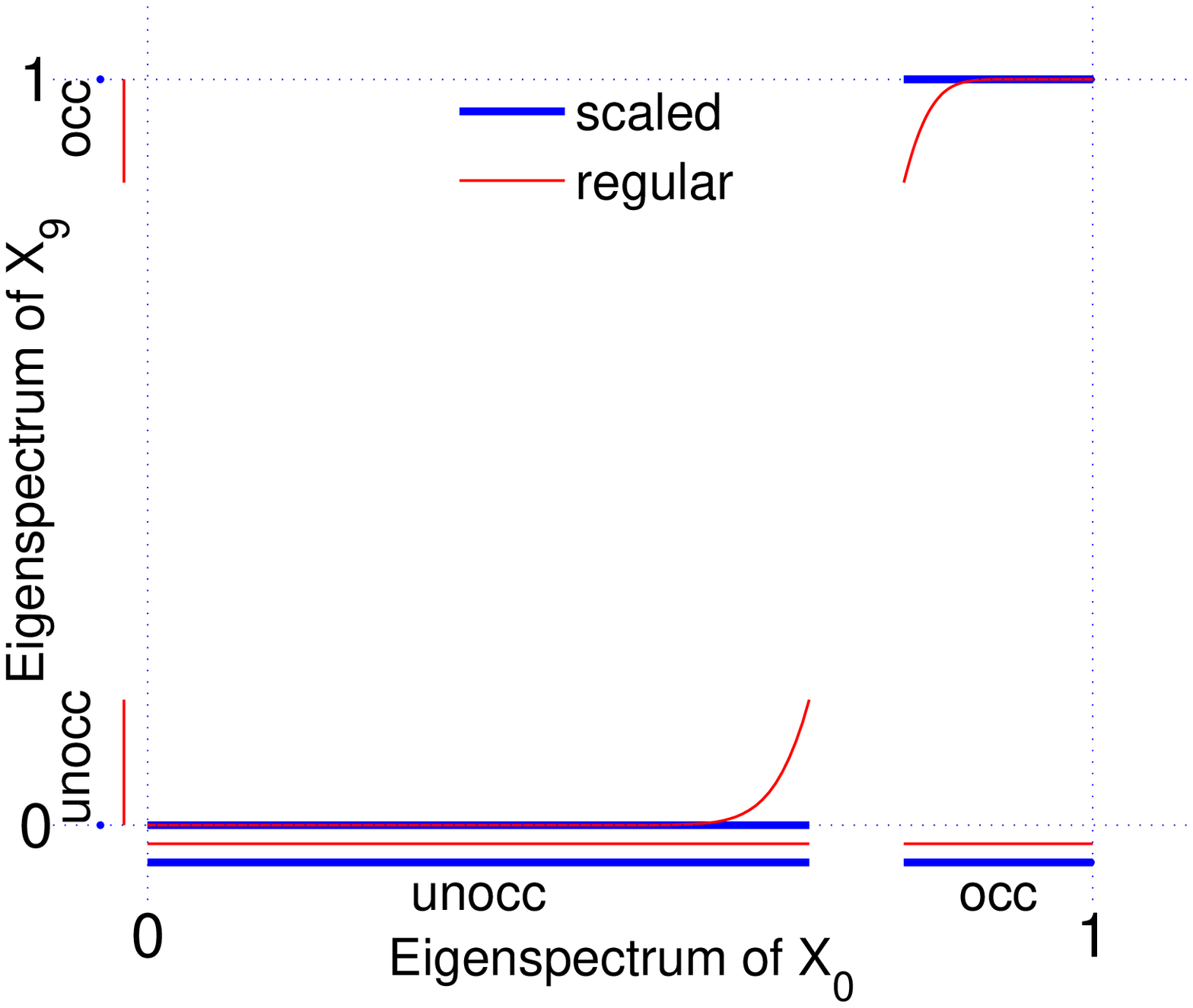}
    }
    \caption{Mapping of the eigenspectrum after 1, 5, and 9 iterations
      respectively of $P_{0,1}$ \& $P_{1,0}$ based varying-$\mu$
      purification with and without use of scaling. In this
      illustrative example $\Delta\epsilon / \xi = 10$ and the
      chemical potential $\mu$ is located at $\lambda_{\textrm{min}} +
      0.25 (\lambda_{\textrm{max}} -
      \lambda_{\textrm{min}})$.  \label{fig:mapping_tc2}}
  \end{figure*}

  The scale and fold technique can also be used together with
  varying-$\mu$ purification. We shall here focus on purification
  based on the polynomials $P_{0,1}$ and $P_{1,0}$. These polynomials
  can be used to adjust the occupation count;\cite{pur-n} if the
  occupation is too high, the $P_{1,0}$ polynomial is applied,
  otherwise $P_{0,1}$ is applied. The scaling should in this case be
  chosen to stretch out the eigenspectrum below $0$ before application
  of $x^2$ and above $1$ before application of $2x-x^2$. The
  purification transformations are
  \begin{equation}
    f_i(X_{i-1}) = P_{1,0}((1+\alpha)X_{i-1} - \alpha I)
  \end{equation}
  and
  \begin{equation}
    f_i(X_{i-1}) = P_{0,1}((1+\alpha)X_{i-1})
  \end{equation}
  where $\alpha \geq 0$ determines the amount of scaling.  A complete
  algorithm is given in Algorithm~\ref{alg:varying_mu_puri_tc2}, where
  $\lambda_{\textrm{lumo}}$ and $\lambda_{\textrm{homo}}$ are the
  eigenvalues closest above and below the band gap, respectively.
  Without scaling, i.e. $\alpha = 0$, this algorithm is essentially
  equivalent to the second order trace correcting purification scheme
  by Niklasson,\cite{pur-n} the only difference being how to choose
  polynomial in line~5 of the algorithm. In the original work by
  Niklasson, the choice was based on the trace of the current density
  matrix approximation. Here, the polynomial is chosen based on the
  eigenvalues $\beta$ and $\bar{\beta}$ that correspond to the lowest
  unoccupied and highest occupied molecular orbitals,
  respectively.\cite{m-accPuri} The behavior of
  Algorithm~\ref{alg:varying_mu_puri_tc2} is illustrated in
  Figure~\ref{fig:mapping_tc2}. The regular scheme with $\alpha = 0$
  is shown for reference.

  \begin{figure*}%[h]
    \center
    \subfigure[$\, $ Varying $\mu$\label{fig:chempot_vs_nmul}]{
      \includegraphics[width=0.4\textwidth]{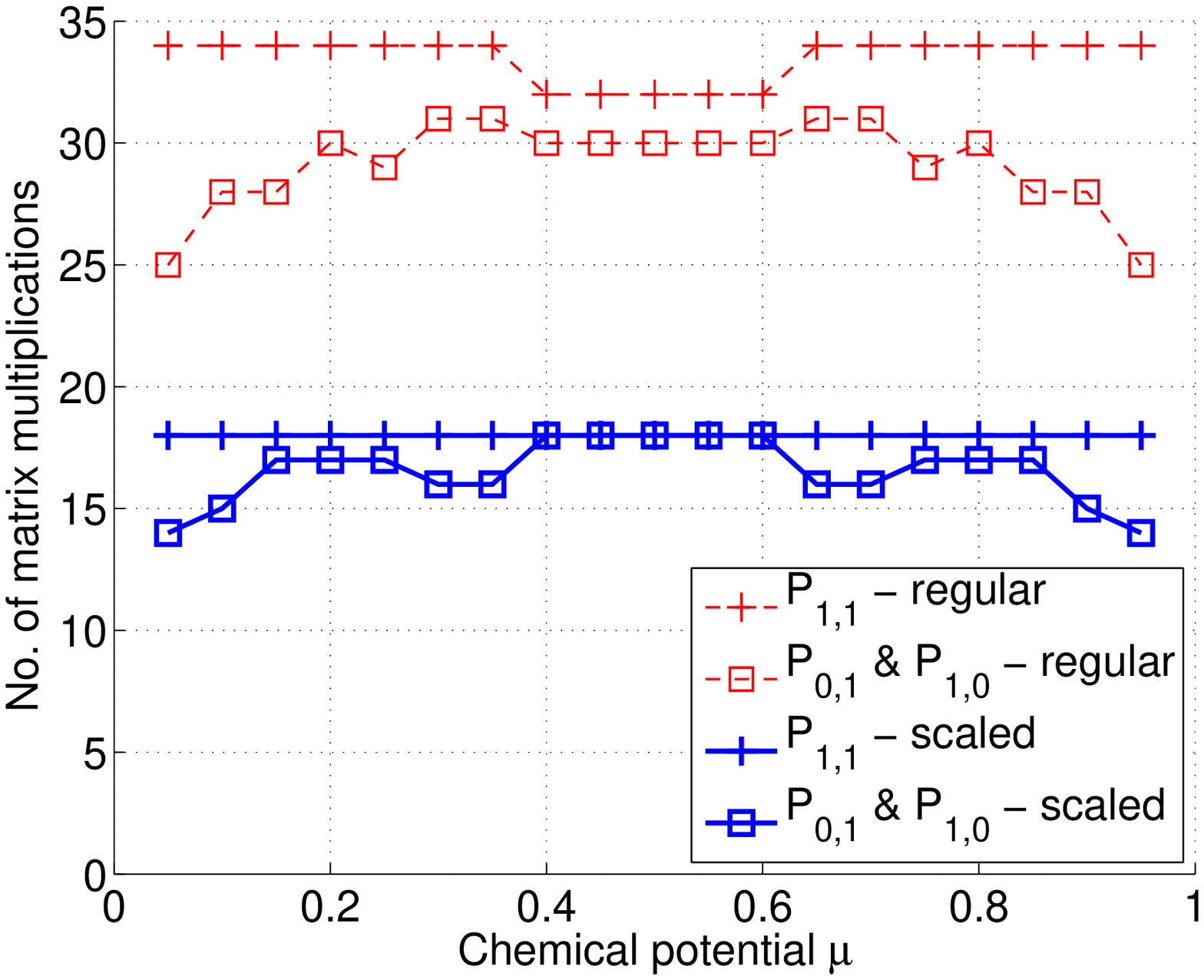}
    }
    \subfigure[$\, $ Varying $\xi$\label{fig:bandgap_vs_nmul}]{
      \includegraphics[width=0.4\textwidth]{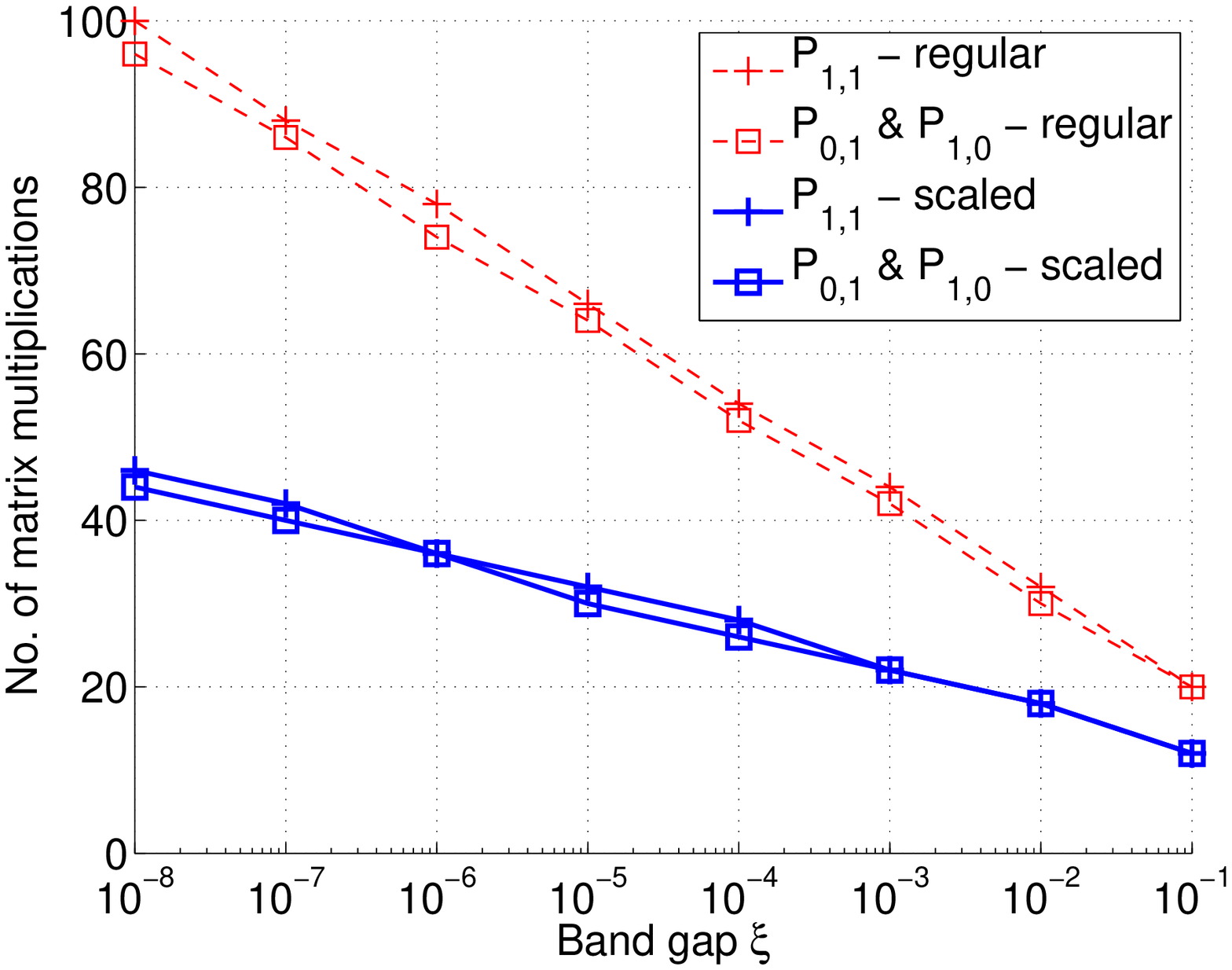}
    }
    \caption{Number of matrix-matrix multiplications needed to reach
      an accuracy of $\|\widetilde{D}-D\|_2 \leq 10^{-9}$, where
      $\widetilde{D}$ is the computed approximation of the exact
      density matrix $D$.  The test calculations presented in
      Panel~(a) were performed on test Hamiltonians with band gaps
      $\xi=0.01$ and varying chemical potential $\mu$. The test
      calculations presented in Panel~(b) were performed on test
      Hamiltonians with chemical potentials $\mu=0.5$ and varying band
      gap $\xi$.  In all cases, the spectral widths of the test
      Hamiltonians were $\Delta\epsilon = 1$.
      The test cases in Panel~(a) are essentially equivalent to the
      test cases presented in Figure~2 of Ref.~\onlinecite{pur-n}.
      \label{fig:performance}}
  \end{figure*}

  Figures~\ref{fig:mapping_mcw} and~\ref{fig:mapping_tc2} show that
  the use of scaling results in more rapid convergence. In order to
  closer study the performance enhancement given by the scaling
  technique we shall consider diagonal test Hamiltonians with varying
  chemical potential and band gap. As previously discussed by
  Maziotti,\cite{pur-m} the results for a given chemical potential and
  a given band gap are valid for any Hamiltonian with that band gap
  and chemical potential.  

  Figure~\ref{fig:chempot_vs_nmul} shows that the proposed scaling
  techniques give significant speedup independently of the location of
  the chemical potential.  As can be seen in
  Figure~\ref{fig:bandgap_vs_nmul}, the cost of the scaled
  purification schemes scale as $\mathcal{O}(1 / \ln \xi )$ with the
  band gap $\xi$, just as for the regular schemes.  However, the
  convergence for the scaled schemes is around twice as fast as for
  the regular schemes.

The scaling technique requires some information about the location of
the band gap. More precisely, a lower bound of the lower edge and an
upper bound of the upper edge of the band gap are needed.  It should
be noted that incorrect bounds can lead to a mix-up between occupied
and unoccupied states. However, even if the bounds are not tight, the
scaling technique can be used although the effect will not be as good
as it could have been.  Tight bounds can be obtained by some technique
for calculation of interior eigenvalues.\cite{interior-v, intEigs,
  m-accPuri}

The performance was here measured by the number of matrix-matrix
multiplications needed to reach a certain accuracy. In practical
linear scaling calculations, efficient ways to bring about sparsity is
critical for the performance.  Since the proposed schemes are on the
standard form given by \eqref{eq:puri_general}, it is possible to
combine them with previously suggested schemes for control of the
forward error.\cite{m-accPuri} As fewer iterations are needed, more
aggressive truncation of small matrix elements can be used in each
iteration.  Therefore, we expect that the speedup given by the
proposed techniques will be even better when the additional problem of
bringing about sparsity is taken into account, although this is
something that needs to be further investigated.

In this letter, non-monotonic recursive polynomial expansions for
calculation of the density matrix were proposed.  We have withdrawn
from the idea that the approximation of the step function should be
monotonically increasing and show that this makes it possible to find
new, more efficient non-monotonic purification transformations.  The
scaled purification variants of this work represent a substantial
improvement compared to previous purification schemes. The reduction
in computational cost is essentially independent of the location of
the chemical potential and the proposed schemes are particularly
efficient in case of small band gaps.

Comments from Sara Zahedi and support from the Swedish Research
Council under Grant No. 623-2009-803 are gratefully acknowledged.

\bibliography{biblio} \bibliographystyle{apsrev} % achemso, apsrev

\begin{thebibliography}{28}
\bibitem[{mhn()}]{mhnm-polys}
\bibinfo{note}{The McWeeny polynomial\cite{hist-mcweeny} is $P_{1,1}(x) =
  3x^2-2x^3$. This polynomial is equivalent to the Newton-Schulz iteration
  polynomial $\frac{1}{2}x(3-x^2)$ for sign matrix
  evaluation.\cite{book-higham} The polynomials suggested by Holas\cite{pur-h}
  can be written in the form $P_{m,m}(x)$. Niklasson\cite{pur-n} proposed
  purification schemes based on polynomials $P_{1,m}(x)$ and $P_{m,1}(x)$.
  Maziotti\cite{pur-m} suggested use of asymmetric polynomials $P_{m,m+1}(x)$
  and $P_{m+1,m}(x)$.}
\end{thebibliography}

\end{document}